\documentclass[aip,apl,reprint,final]{revtex4-2}

\usepackage[colorlinks=true,allcolors=blue]{hyperref} 
\usepackage{graphicx}
\usepackage{amsmath,amssymb,mathrsfs,array,longtable,delarray}
\usepackage{latexsym}
\usepackage{dcolumn}
\usepackage{bm}
\usepackage{textcomp} 
\usepackage{siunitx}
\usepackage{verbatim}

\begin{document}

\title{Field effect two-dimensional electron gases in modulation-doped InSb surface quantum wells}

\author{E. Annelise Bergeron}
\affiliation{Institute for Quantum Computing, University of Waterloo, Waterloo N2L 3G1, Canada}
\affiliation{Department of Physics, University of Waterloo, Waterloo N2L 3G1, Canada}

\author{F. Sfigakis}
\altaffiliation{corresponding author: francois.sfigakis@uwaterloo.ca}
\affiliation{Institute for Quantum Computing, University of Waterloo, Waterloo N2L 3G1, Canada}
\affiliation{Northern Quantum Lights inc., Waterloo N2B 1N5, Canada}
\affiliation{Department of Chemistry, University of Waterloo, Waterloo N2L 3G1, Canada}

\author{Y. Shi}
\affiliation{Department of Electrical and Computer Engineering, University of Waterloo, Waterloo N2L 3G1, Canada}
\affiliation{Department of Physics, University of Waterloo, Waterloo N2L 3G1, Canada}
\affiliation{Waterloo Institute for Nanotechnology, University of Waterloo, Waterloo N2L 3G1, Canada}

\author{George Nichols}
\affiliation{Institute for Quantum Computing, University of Waterloo, Waterloo N2L 3G1, Canada}
\affiliation{Department of Physics, University of Waterloo, Waterloo N2L 3G1, Canada}

\author{P. C. Klipstein}
\affiliation{Semiconductor Devices, P.O. Box 2250, Haifa 31021, Israel}

\author{A. Elbaroudy}
\affiliation{Department of Electrical and Computer Engineering, University of Waterloo, Waterloo N2L 3G1, Canada}
\affiliation{Department of Physics, University of Waterloo, Waterloo N2L 3G1, Canada}

\author{Sean M. Walker}
\affiliation{Institute for Quantum Computing, University of Waterloo, Waterloo N2L 3G1, Canada}
\affiliation{Department of Chemistry, University of Waterloo, Waterloo N2L 3G1, Canada}

\author{Z. R. Wasilewski}
\affiliation{Institute for Quantum Computing, University of Waterloo, Waterloo N2L 3G1, Canada}
\affiliation{Department of Physics, University of Waterloo, Waterloo N2L 3G1, Canada}
\affiliation{Northern Quantum Lights inc., Waterloo N2B 1N5, Canada}
\affiliation{Department of Electrical and Computer Engineering, University of Waterloo, Waterloo N2L 3G1, Canada}
\affiliation{Waterloo Institute for Nanotechnology, University of Waterloo, Waterloo N2L 3G1, Canada}

\author{J. Baugh}
\altaffiliation{baugh@uwaterloo.ca}
\affiliation{Institute for Quantum Computing, University of Waterloo, Waterloo N2L 3G1, Canada}
\affiliation{Department of Physics, University of Waterloo, Waterloo N2L 3G1, Canada}
\affiliation{Northern Quantum Lights inc., Waterloo N2B 1N5, Canada}
\affiliation{Department of Chemistry, University of Waterloo, Waterloo N2L 3G1, Canada}
\affiliation{Waterloo Institute for Nanotechnology, University of Waterloo, Waterloo N2L 3G1, Canada}

\begin{abstract}
We report on transport characteristics of field effect two-dimensional electron gases (2DEG) in surface indium antimonide quantum wells. The topmost 5 nm of the 30 nm wide quantum well
is doped and shown to promote the formation of reliable, low resistance Ohmic contacts to surface InSb 2DEGs. High quality single-subband magnetotransport with clear quantized integer quantum Hall plateaus are observed to filling factor $\nu=1$ in magnetic fields of up to $B=18$ T. We show that the electron density is gate-tunable, reproducible, and stable from pinch-off to 4$\times 10^{11}$ cm$^{-2}$, and peak mobilities exceed 24,000 cm$^2$/Vs. Large Rashba spin-orbit coefficients up to 110 meV$\cdot${\AA} are obtained through weak anti-localization measurements. An effective mass of 0.019$m_e$ is determined from temperature-dependent magnetoresistance measurements, and a g-factor of 41 at a density of 3.6$\times 10^{11}$ cm$^{-2}$ is obtained from coincidence measurements in tilted magnetic fields. By comparing two heterostructures with and without a delta-doped layer beneath the quantum well, we find that the carrier density is stable with time when doping in the ternary Al$_{0.1}$In$_{0.9}$Sb barrier is not present. Finally, the effect of modulation doping on structural asymmetry between the two heterostructures is characterized.
\end{abstract}

\maketitle

Confining potentials in electrostatically-defined nanoscale devices, such as single electron transistors or single electron pumps, are strongly enhanced in two-dimensional electron gases (2DEGs) hosted at the surface or near the surface of semiconductor heterostructures.\cite{buonacorsi2021non} Furthermore, surface or near-surface quantum well (QW) heterostructures in III-V semiconductors are compatible with proximitized superconductivity and offer a scalable planar platform for superconductor-semiconductor systems, such as those suggested for topological quantum computation\cite{shabani2016two,karzig2017scalable} and
those suitable for topological phase transitions involving Majorana zero modes.\cite{moore2018quantized,lee2019contribution,ke2019ballistic} Amongst III-V binary semiconductors, Indium Antimonide (InSb) has the smallest electron effective mass, highest spin orbit coupling \cite{khodaparast2004spectroscopy, khodaparast2004spin}, and largest Land\'e g-factor. Such material properties makes the pursuit of InSb QWs desirable for a number of quantum device applications, including quantum sensing, quantum metrology, and quantum computing.

High quality two-dimensional electron gases (2DEGs) in InSb QWs are difficult to realize partly due to the highly mismatched lattice constants between the quantum well and barrier materials,\cite{lehner2018limiting} the available purity of the required base elements (In, Sb),\cite{note02} and the lack of wafer-to-wafer reproducibility with doping schemes.\cite{lehnerPhD} InSb QWs have generally relied on the use of modulation-doping for 2DEG formation, but these structures have frequently reported issues with parasitic parallel conduction and unstable carrier densities.\cite{lei2019quantum,qu2016quantized,kulesh2020quantum,lei2021gate} This is especially true of InSb surface QWs, which must contend with a Schottky barrier at the surface. Dopant-free field-effect 2DEGs avoid these issues and have recently been reported in undoped InSb QWs.\cite{lei2022high} However, as reported in GaAs systems, achieving good Ohmic contacts is challenging in completely undoped heterostructures, especially near the surface.\cite{mondal2014field}

In this Letter, we report on the use of a thin $n$-InSb layer to promote the formation of reliable, low resistance Ohmic contacts to a surface InSb QW. We compare two InSb surface QW heterostructures, one with and one without a delta-doped Al$_{0.1}$In$_{0.9}$Sb layer, and demonstrate the influence of modulation doping on gating characteristics, magnetotransport behavior, and spin-orbit interaction. We overcome issues of parallel conduction in both heterostructures and present magnetotransport behavior of a high quality, single-subband 2DEG up to 18 T. The effective mass, transport and quantum lifetimes, and g-factor are determined from magnetoresistance measurements. The strength of Rashba spin-orbit interaction is characterized using weak anti-localization measurements.

\begin{figure}[h]
  \includegraphics[scale=0.95]{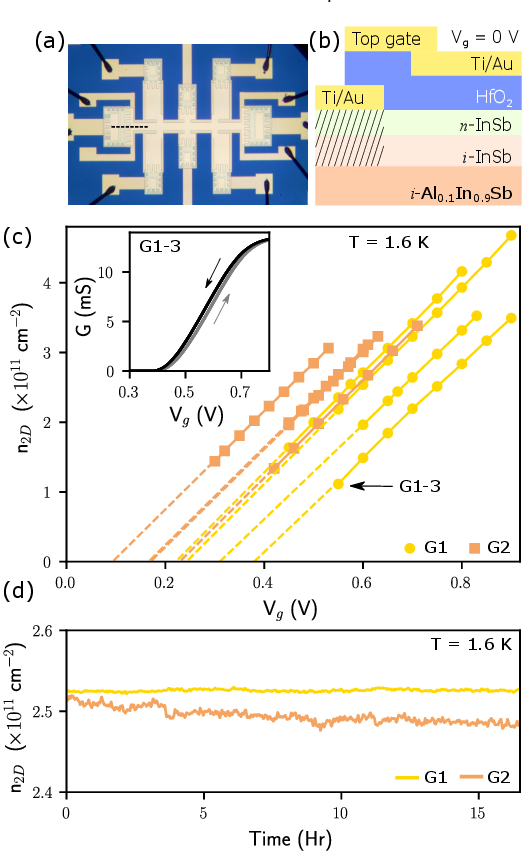}
  \caption{\label{Gate} (a) Optical image of a representative gated Hall bar. The global top-gate overlaps the Ohmic contacts in order to induce a 2DEG between contacts. (b) Schematic of the cross section along the dotted line in (a). The 30 nm InSb quantum well is populated by electrons beneath Ti/Au Ohmic contacts (hashed region), unlike regions directly underneath HfO$_2$ (see main text). (c) Hall density versus top-gate voltage of all eight Hall bars from G1 and G2. The 2DEG density increases linearly with $V_g$ in all samples, and is reproducible along the linear traces. (inset) Two-terminal differential conductance $G(V_g)=dI/dV_{sd}$ (using 100 \textmu V ac excitation) showing the turn-on voltage of a gated Hall bar on G1. Eight traces are shown, four while increasing $V_g$ (grey) and four while decreasing $V_g$ (black). (d) The Hall density in G1 remains stable for 16 hours, whereas it drifts with time in G2.}
\end{figure}

Two wafers, G1 and G2, were grown by molecular beam epitaxy (MBE). Wafer G1 had the following sequence of layers (see Figure \ref{Gate}), starting from a 3’’ semi-insulating (SI) GaAs (001) substrate: a 120 nm GaAs smoothing layer, 100 nm AlSb nucleation layer, a 4 \textmu m Al$_{0.1}$In$_{0.9}$Sb dislocation filter buffer, a 30 nm InSb quantum well where the topmost 5 nm was doped with Si at a doping density of $2 \times 10^{18}$ cm$^{-3}$. Wafer G2 is identical to wafer G1, except for an additional Si delta-doped layer (with sheet doping density $1.5 \times 10^{11}$ cm$^{-2}$) located 10 nm below the InSb quantum well. Section I in the Supplementary Material provides additional details about MBE growth. In both wafers, the doped $n$-InSb layer facilitates the low-temperature formation of low-resistance Ohmic contacts to the 2DEG. The purpose of the delta-doped layer below the InSb quantum well in G2 is to pull the 2DEG wavefunction further away from the surface than in G1. Section II from the Supplementary Material shows self-consistent simulations\cite{NextNano-A,NextNano-B,NextNano-C} of the bandstructure profiles for both G1 and G2.

\begin{table}[t]
    \begin{ruledtabular}
    \begin{tabular}{cccc}
    \multicolumn{2}{c}{Wafer G1} & \multicolumn{2}{c}{Wafer G2} \vspace{0.5 mm} \\
    \cline{1-2} \cline{3-4} \vspace{-2.5 mm} \\
    Hall bar & peak mobility & Hall bar & peak mobility \\
    ID & (cm$^2$/Vs) & ID & (cm$^2$/Vs) \vspace{0.5 mm} \\ \hline \vspace{-2.5 mm} \\
    G1-1 & 18,000 & G2-1 & 21,800 \\
    G1-2 & 22,200 & G2-2 & 24,400 \\
    G1-3 & 21,200 & G2-3 & 24,600 \\
    G1-4 & 23,100 & G2-4 & 24,100
    \end{tabular}
    \end{ruledtabular}
    \caption{List of all samples reported in this Letter.}
    \label{tab:hallbars}
\end{table}

Eight gated Hall bars (see Table \ref{tab:hallbars} and Figure \ref{Gate}) were fabricated using standard optical lithography and wet-etching techniques, keeping all processes at or below a temperature of 150$^\circ$C to prevent the deterioration of device characteristics,\cite{uddin2013gate,yi2015gate,kulesh2020quantum} and preventing the InSb surface from coming into contact with photoresist developer (see Section III of the Supplementary Material for more details on sample fabrication). Ti/Au Ohmic contacts were deposited directly on the doped $n$-InSb layer. Immediately prior to deposition, Ohmic contacts were treated with a sulfur passivation solution. The latter is designed to etch away native oxides, prevent further surface oxidation during transfer in air to the deposition chamber, and possibly dope the surface.\cite{tajik2012photoluminescence,lebedev2020modification,bessolov1998chalcogenide} Combined with the presence of Si dopants at the surface of the InSb quantum well, Ohmic contacts with typical resistances of 400$-$800 $\Omega$ were achieved in zero magnetic field, and $\sim$\,12~k$\Omega$ at $B=18$ T. Finally, 60 nm thick hafnium dioxide (HfO$_2$) was deposited by atomic layer deposition (ALD) at 150$^\circ$C, followed by the deposition of a Ti/Au global top-gate that overlaps the Ohmic contacts.

Using standard ac lock-in measurement techniques (see Section IV of the Supplementary Material for circuit diagrams and experimental details), four-terminal and two-terminal transport experiments were performed in a pumped-$^4$He cryostat and a $^3$He/$^4$He dilution refrigerator, with a base temperature of 1.6 K and 11 mK respectively. In ungated Hall bars, the as-grown electron densities of G1 and G2 were $3.0 \times 10^{11}$ cm$^{-2}$ and $3.5 \times 10^{11}$ cm$^{-2}$, respectively. However, in all gated Hall bars, the quantum well in both wafers is completely depleted of electrons at top-gate voltage V$_g=0$, most likely due to significant trapped charges associated with HfO$_2$.\cite{yi2015gate,Baik2017} A positive top-gate voltage is needed for a 2DEG to form. The 2DEG turn-on threshold voltage is the intercept of the electron density n$_{2D}$(V$_g$) on the top-gate voltage axis in Figure \ref{Gate}(c), obtained from the linear extrapolation of the data for each Hall bar to n$_{2D}=0$.\footnote{This definition removes any ambiguity in the turn-on threshold due to the transition from the Boltzmann transport regime to the percolation regime at low electron densities.} The average 2DEG turn-on threshold is V$_g=(0.29 \pm 0.06)$ V for wafer G1 and V$_g=(0.17 \pm 0.05)$ V for wafer G2. The lower threshold in wafer G2 is consistent with the additional doping provided by its delta-doped layer, which brings the conduction band closer to the Fermi level in wafer G2 than in wafer G1. Magnetotransport plots at $T=1.6$ K for six samples are shown in Section IV of the Supplementary Material.

The inset of Figure \ref{Gate}(c) shows a typical pinch-off curve for a gated Hall bar from wafer G1 in a two-terminal conductance measurement. Agreement between the  pinch-off voltage ($V_g=0.38$ V) from the two-terminal measurement and the extrapolated 2DEG turn-on threshold ($V_g=0.38$ V) from the four-terminal measurement, both obtained from the same Hall bar, strongly indicates that there is no significant tunnel barrier within the Ohmic contacts themselves.\cite{FujitaT21} Indeed, the electron density in the InSb quantum well directly underneath the Ohmic contact metal should be the same as or very similar to the as-grown electron density, because the HfO$_2$ dielectric is not in direct contact with $n$-InSb (\textit{i.e.}, there is not a large trapped charge density). The pinch-off curves are stable and reproducible, overlapping perfectly when $V_g$ is swept in the same direction and showing minimal hysteresis when $V_g$ is swept in opposite directions. After pinch-off, the 2DEG does not turn itself back on with time.\cite{lei2022high,qu2016quantized,lei2021gate} To further illustrate this time stability, Figure \ref{Gate}(d) shows the carrier density measured over a period of 16 hours, where it essentially stays constant. This is not however the case with devices from G2, where the electron density can drift with time. We speculate this could be due to the presence of dopants in the Al$_{0.1}$In$_{0.9}$Sb layer. Indeed, quantum dots fabricated in InSb 2DEGs with modulation-doped AlInSb have recently been reported where their characteristics drift in time.\cite{kulesh2020quantum}

\begin{figure}[h]
  \includegraphics[scale=0.90]{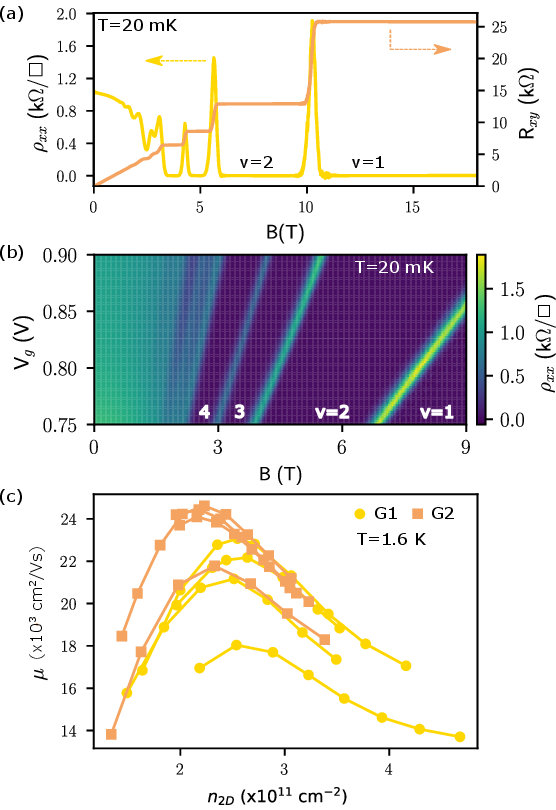}
  \caption{\label{Transport} (a) Longitudinal resistivity $\rho_{xx}$ and Hall resistance $R_{xy}$ vs. magnetic field at $n_{2D}=3.4\times$10$^{11}$ cm$^{-2}$. (b) Landau fan diagram. Integer quantum Hall states from $\nu = 1$ to 4 are labeled. (c) Mobility vs. Hall density of all Hall bars from G1 (circles) and G2 (squares).}
\end{figure}

Figure \ref{Transport}(a) shows the transverse (Hall) resistance $R_{xy}$ and longitudinal resistivity $\rho_{xx}$ in a magnetic field up to $B=18$ T at the highest accessible carrier density $3.4 \times 10^{11}$ cm$^{-2}$ for sample G1-3. The transverse resistance exhibits well-defined quantized quantum Hall plateaus $R_{xy}=h/\nu e^2$ at filling factors $\nu = hn_{2D}/eB = 1, 2, 3,$ and 4, where $h$ is the Planck constant and $e$ is the single electron charge. The population of a single subband is evidenced by the observation of single-frequency Shubnikov-de-Haas oscillations in combination with vanishing $\rho_{xx}=0$ at $\nu = 1, 2, 3, 4$. Furthermore, the 2DEG density determined from the periodicity of Shubnikov-de-Haas oscillations versus inverse field, given by $n_{2D} = \frac{2e}{h}\left( \frac{1}{B_{\nu+1}} - \frac{1}{B_{\nu}}\right)^{-1}$, matches the total carrier density $n_{tot}$ determined via the classical Hall effect $n_{tot}=B/eR_{xy}$. No signs of parallel conduction from either a second subband or another conductive layer is discernible. The absence of Landau level crossings in the Landau fan diagram shown in Figure \ref{Transport}(b) indicates the single subband behavior persists over the entire measured density range. The Landau fan, obtained by sweeping the top-gate at magnetic field increments on sample G1\nobreakdash-3, showcases the reproducibility and stability of gating characteristics.

The dependence of the transport mobility $\mu$ on 2DEG density is shown in Figure \ref{Transport}(c) which shows an average peak mobility of $(2.1\pm 0.2) \times 10^{4}$ cm$^{2}$/Vs near $n_{2D} = 2.5 \times 10^{11}$ cm$^{-2}$ in G1. The decrease in mobility at higher densities is attributed to increasing interface roughness scattering\cite{ando1982electronic,arjun2022} as the electron wavefunction is pulled closer to the surface by the increasing electric field of the top-gate. Increased scattering from a populating second subband is ruled out, since there is only one subband populated over that range of density. Alloy scattering (typically only observed in ternary alloys) is also ruled out, since the 2DEG wavefunction lies almost entirely within the InSb quantum well. The higher average peak mobility of $(2.4 \pm 0.1) \times 10^{4}$ cm$^2$/Vs near $n_{2D} = 2.2 \times 10^{11}$ cm$^{-2}$ in G2 is consistent with its 2DEG being pulled further away from the surface by the delta-doped layer, relative to G1. The greater device-to-device reproducibility in G2 than in G1 is also consistent with this picture. Variability between nominally identical devices may be mostly due to surface treatment during sample fabrication. The mobilities reported here could perhaps be improved further\cite{lehner2018limiting} by reducing the density of threading dislocations\cite{shi2017, mishima2005effect} and hillocks.\cite{shi2019, chung2000improving}

\begin{figure}[t]
  \includegraphics[scale=0.90]{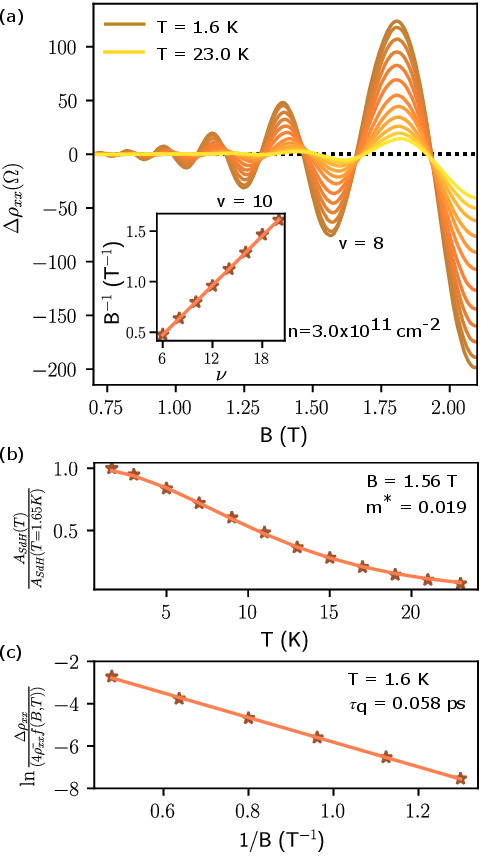}
  \caption{\label{Temp} (a) Temperature-dependent amplitudes of SdH oscillations at fixed  $n_{2D}=\num{3e11}$ cm$^{-2}$ in G2 where $\Delta\rho_{xx}$ is obtained by subtracting a polynomial background from $\rho_{xx}$. (inset) The 1/B values of the minima in $\rho_{xx}$ are plotted versus $\nu$. The 2DEG density, determined from the periodicity of SdH oscillations, is given by the slope of the line $|e|/hn_{2D}$. (b) The temperature dependent amplitude of the $\nu=8$ minima at $B=1.56$ T in (a), normalized by its value at $T=1.6$ K. The line is a fit to a temperature dependent factor, discussed in the Supplementary Material, to determine the effective mass. A value of m$^*=0.0189 \pm 0.0001$ at 1.56 T is found for a 2DEG density of $\num{3e11}$ cm$^{-2}$. (d) The effective mass is used to determine the the quantum lifetime from a Dingle plot given by $\ln({\Delta\rho_{xx}/4\bar{\rho}_{xx}f(B,T)})$ vs. inverse magentic field. Data points corresponding to the minima in the oscillations of the $T=1.6$ K trace in (a) are plotted versus $1/B$. A quantum lifetime of 0.58 ps is determined from the slope of the resulting straight line $-\pi m^*/|e|\tau_q$. }
\end{figure}

Figure \ref{Temp}(a) shows the temperature dependence of the amplitude of low-field SdH oscillations $\Delta\rho_{xx}$ in sample G2-4, obtained by subtracting a polynomial background from $\rho_{xx}$. The data was taken at a density of \num{3e11} cm$^{-2}$, determined from the periodicity of SdH oscillations versus inverse magnetic field shown in the inset. The temperature-dependent amplitude A$_{SdH} (T)$ of the $\nu=8$ minimum at $B=1.56$ T, normalized by the base temperature value A$_{SdH}$ ($T=1.6$ K), is plotted in Figure \ref{Temp}(b), and fit to theory (see Section V of the Supplementary Material for more details). A value of $m^* = (0.0189\pm 0.0001)m_e$ is obtained, which is higher than 0.014$m_e$ found in bulk InSb. This larger value for the QW is found to agree quite well with the predictions of an 8-band  $\mathbf{k\cdot p}$ calculation for a symmetric InSb/In$_{0.9}$Al$_{0.1}$Sb QW, as presented in Section VI of the Supplementary Material. Although our QW is not symmetrical, the contribution due to wave function penetration of the barrier layers is shown to be quite small. The most dominant contributions to the mass increase appear to come from enlargement of the QW band gap due to confinement and strain, and from the strong non-parabolicity of the electron dispersion.  It should be noted that our experimental fit gives an average parabolic mass that matches the number of states in the filled Landau levels to the number of states in the real non-parabolic dispersion.

Using the $T=1.6$ K trace in Figure \ref{Temp}(a), a quantum lifetime $\tau_q=0.058$ ps, also known as the single-particle relaxation time, is extracted from the Dingle plot shown in Figure \ref{Temp}(c) (see Section V of the Supplementary Material for more details). In comparison, the mean transport lifetime derived from the Drude model $\tau_t = \mu m^*/e$ is 0.21 ps. The ratio of transport to quantum lifetimes is thus $\tau_t/\tau_q \approx 4$.  Since $\tau_t$ is weighted by the scattering angle whereas $\tau_q$ is related to total scattering, the ratio $\tau_t/\tau_q$ provides insight into the nature of scattering affecting transport.\cite{coleridge1989low} For transport mobilities limited by large angle scattering (as is the case here due to interface roughness), the ratio approaches unity.  In other binary QW heterostructures, large ratios of $\sim$ 40 have been reported in samples where small angle scattering from long range potentials (e.g., remote ionized impurities) was the dominant scattering mechanism, leading to high mobilities and long transport lifetimes \cite{lee2019contribution}. Although our transport lifetime differs by more than a factor of ten from these reports, the quantum lifetimes are comparable and justify our use of dopants in the QW.

\begin{figure}[t]
  \includegraphics[width=0.95\linewidth]{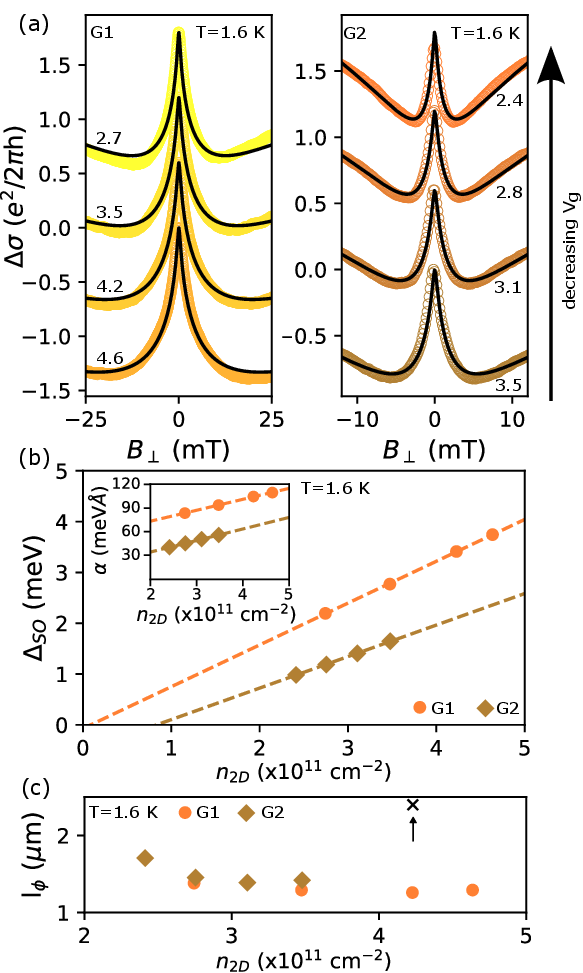}
  \caption{\label{WAL} (a) Density dependence of weak anti-localization in G1 (left) and G2 (right). Experimental points are displayed as colored open circles and fits to the HLN model are shown as black lines. Curves are offset for clarity and labeled with the corresponding 2DEG density in units of $1\times 10^{11}$ cm$^{-2}$. (b) Spin orbit splitting $\Delta_{so}$ vs density extracted from HLN fits to data in (a). A linear increase in $\Delta_{so}$ is observed with increasing density in both wafers. (inset) Spin orbit coeficient $\alpha_{so} = \Delta_{so}/2k_F$ as a function of $n_{2D}$. (c) Phase coherence length $l_{\phi}$ vs. density acquired from HLN fits to data in (a). Also shown and indicated by a black arrow is the phase coherence in G1 measured at a temperature of 22 mK. }
\end{figure}

The Land\'{e} g-factor $g^*$ was measured in sample G1\nobreakdash-3 at $\nu = 4$ for two different carrier densities, using a tilted magnetic field approach\cite{Fang68tilted} to identify coincidences between the Zeeman and cyclotron energies $g^*\mu_BB_{\text{tot}}=\hbar eB_\perp/m^*$, where $\mu_B$ is the Bohr magneton, $B_{\text{tot}}$ is the total magnetic field, and $B_\perp$ is the component of $B_{\text{tot}}$ that is perpendicular to the 2DEG plane. By modeling the evolution of spin-split Landau energy levels, the effective g-factors $g^*=33\pm2$ at $\num{2.8e11}$ cm$^{-2}$ and $g^*=41\pm2$ at $\num{3.6e11}$ cm$^{-2}$ were obtained, in  agreement with other reports of the effective g-factor in InSb.\cite{lei2020electronic,lei2022high,Nedniyom09giant,Yang11tilted,moehle2021} Section VII in the Supplementary Material contains more experimental and theoretical details of the coincidence experiments.

Wafers G1 and G2 are characterized by a strong and tunable spin orbit interaction (SOI), as demonstrated by the weak anti-localization (WAL) conductivity peak observed in all Hall bars. The strength of SOI was determined from fits to $\Delta\sigma_{xx}$ using the Hikami-Larkin-Nagaoka (HLN) model,\cite{hikami1980spin} where $\Delta\sigma_{xx} = \sigma_{xx}(B)-\sigma_{xx}(0)$, $\sigma_{xx}(B)$ is the field-dependent conductivity, and $\sigma_{xx}(0)$ is a constant background conductivity. Figure \ref{WAL}(b) shows the density dependence of the spin-orbit strength in samples G1-1 and G2-1, obtained from fits presented in Figure \ref{WAL}(a). The Rashba coefficient $\alpha_{so}$ is related to spin orbit length via $\alpha_{so} = \Delta_{so}/2k_F$ and  $\Delta_{so} = \sqrt{2\hbar^2/\tau_D \tau_{so}}$ where $\Delta_{so}$ is the energy gap, $k_F$ is the Fermi wave vector, $\tau_D$ is the diffusion time, and $\tau_{so}$ is the spin orbit time. The Rashba coefficient $\alpha_{so}$ reaches a maximum of nearly 110 meV$\cdot\text{\AA}$ at $n_{2D} = 4.6 \times 10^{11}$ cm$^{-2}$ in G1, among the highest values reported in the literature for InSb\cite{gilbertson2009zero,khodaparast2004spectroscopy,lei2022arxiv,lei2022high,kallaher2010,moehle2021} (see Table S2 from Section VIII in the Supplementary Material for an explicit comparison), but not as high as in InSbAs.\cite{moehle2021,metti2022} Being related to structural asymmetry, $\alpha_{so}$ is enhanced by the asymmetry of the wavefunction in the QW at high electric fields. Three factors significantly enhance structural asymmetry in G1: (i) the \textit{n}-InSb layer inside the QW, (ii) the high-$\kappa$ dielectric HfO$_2$ (with $\epsilon \approx 20$), and (iii) the 2DEG location at the surface. Comparing wafers G1 and G2 in Figure \ref{WAL}(b), $\alpha_{so}$ is weakened in G2 by nearly a factor of two for all devices measured. The delta-doped Al$_{0.1}$In$_{0.9}$Sb layer in G2 is responsible for this behavior: it causes band bending that pulls the 2DEG wavefunction towards the center of the QW, thereby reducing the structural asymmetry and Rashba component of the SOI (see section II in the Supplementary Material for bandstructure profiles of G1 and G2). There is thus a trade-off between mobility and strength of Rashba interactions. Figure \ref{WAL}(c) shows the phase coherence lengths $l_\phi$ determined from the fits to the HLN model are slightly larger in G2 than those in G1 at $T=1.6$ K. Within the same wafer, $l_\phi$ reaches a maximum at the same density as the peak mobility. The phase coherence in G1 increases to 2.4 \textmu m at 22 mK from 1.5 \textmu m at 1.6 K in the same device at a similar carrier density (see Section VIII of the Supplementary Material). In contrast, $\alpha_{so}$ remains constant from 22 mK to 1.6 K.

In conclusion, we presented the growth, fabrication, and transport characteristics of high-quality, gate-tunable InSb 2DEGs in surface quantum wells grown on (001) SI-GaAs substrates. An $n$-InSb layer within the quantum well was used to realize reliable, low-resistance Ohmic contacts. Magnetoresistance measurements confirmed that intentional dopants in InSb are compatible with high-quality and reproducible transport characteristics, without parasitic parallel conduction or unstable carrier densities. The observation of Rashba coefficients $\alpha_{so}$ among the highest values reported in the literature for InSb validates the approach
of using surface quantum wells. Preliminary evidence suggests intentional dopants in Al$_x$In$_{1-x}$Sb might be responsible for the drift with time of transport characteristics. If correct, future modulation-doped InSb 2DEG heterostructure with time-wise stable transport characteristics could be achieved by using an InSb/Al$_{x}$In$_{1-x}$Sb short-period superlattice (SPSL) doping scheme,\cite{Manfra2014} where only the very thin (6$-$10 monolayers) InSb layers are doped.

\section*{Supplementary Material}

The eight sections in the Supplementary Material contain additional information on MBE growth, bandstructure profiles, sample fabrication, magnetotransport characterization of Hall bars, effective mass and quantum lifetime measurements, $\mathbf{k\cdot p}$ calculations, coincidence experiments and modeling, and weak anti-localization experiments.

E.A.B. and F.S. contributed equally to this paper. The authors thank Christine Nicoll for useful discussions. E.A.B. acknowledges support from a Mike and Ophelia Lazaridis Fellowship. This research was undertaken thanks in part to funding from the Canada First Research Excellence Fund (Transformative Quantum Technologies) and the Natural Sciences and Engineering Research Council (NSERC) of Canada. The University of Waterloo's QNFCF facility was used for this work. This infrastructure would not be possible without the significant contributions of CFREF-TQT, CFI, ISED, the Ontario Ministry of Research and Innovation, and Mike and Ophelia Lazaridis. Their support is gratefully acknowledged.

\providecommand{\noopsort}[1]{}\providecommand{\singleletter}[1]{#1}

\end{document}


\maketitle

\section*{\Large{Supplementary Material}}
\vspace{-3mm}\noindent {\textbf{\large{Field effect two-dimensional electron gases in modulation-doped\\ InSb surface quantum wells}}\vspace{1mm}\\
\noindent \textit{E. Annelise Bergeron, F. Sfigakis, Y. Shi, George Nichols, P. C. Klipstein,}\vspace{-2mm}\\
\noindent \textit{A. Elbaroudy, Sean M. Walker, Z. R. Wasilewski, and J. Baugh}

~\\~\\ \noindent \textbf{Table of Contents}\\~\vspace{-5mm}\\
\indent\qquad Section \ref{sec:MBE}: MBE growth methods \\
\indent\qquad Section \ref{sec:bandstructure}: Bandstructure profiles \\
\indent\qquad Section \ref{sec:fabrication}: Fabrication methods \\
\indent\qquad Section \ref{sec:magnetotransport}: Additional magnetotransport characterization \\
\indent\qquad Section \ref{sec:mass}: Effective mass and quantum lifetime \\
\indent\qquad Section \ref{sec:kdotp}: 8-band  $\mathbf{k\cdot p}$ model of InSb/Al$_{0.1}$In$_{0.9}$Sb quantum wells \\
\indent\qquad Section \ref{sec:gfactor}: Coincidence experiment and g-factor\\
\indent\qquad Section \ref{sec:WAL}: Weak anti-localization and spin-orbit interactions
\\

\section{MBE growth methods}
\label{sec:MBE}

The two surface InSb quantum well (QW) heterostructures were grown on 3” semi-insulating GaAs (001) substrates using a Veeco Gen10 MBE system. A growth rate of around 1.8 {\AA}/s was used for the InSb layers and 2 {\AA}/s was used for the rest of the layers in both structures. The Sb/III flux ratio was kept at about 2$-$2.5, where a ratio of 1 corresponds to the minimum Sb flux necessary to sustain group V stabilized surface reconstruction. The wafer was radiatively heated by the substrate manipulator in the growth chamber with a proportional-integral-derivative (PID) controller and a thermocouple positioned on the back of but not in contact with the substrate. The AlSb and AlInSb metamorphic buffers were grown at a substrate temperature of around 500 $^\circ$C and 380 $^\circ$C respectively. The QW regions, including the undoped and Si bulk-doped InSb layers in both structures as well as the Si delta-doping layer and the Al$_{0.1}$In$_{0.9}$Sb spacer in G2, were grown at around 360 $^\circ$C to accommodate the lower sublimation temperature of InSb and to reduce the effect of Si segregation. From an independent study, the secondary ion mass spectrometry (SIMS) was measured on an InSb QW structure grown at similar conditions containing a 50 nm Si bulk-doped Al$_{0.12}$In$_{0.88}$Sb layer with a high doping density of 5 $\times$ 10$^{18}$ cm$^{-3}$ and no signs of Si segregation was observed in the structure.\cite{peytonthesis} The substrate temperature was closely monitored during the growth with the band-edge spectrometer (BET) in the beginning and was then switched to the integrated spectral pyrometry\cite{AlanTam17} (ISP) as the GaAs absorption edge became progressively undetectable with increasing thickness of the narrow gap Al$_x$In$_{1-x}$Sb buffer. More details of growth procedures on similar structures can also be found in Ref.~\onlinecite{Peyton19}.

\section{Bandstructure profiles}
\label{sec:bandstructure}

Figure \ref{bandstructure} shows bandstructure profiles calculated from self-consistent simulations solving both the Poisson and Schr\"{o}dinger equations.\cite{NextNano-A,NextNano-B,NextNano-C} The only structural difference between G1 and G2 is the delta-doped layer in G2, all other parameters are the same for both wafers.

Three observations can be made. \textit{(i)} At $V_g=0$ ($n_{2D}=0$), Figures \ref{bandstructure}(a) and \ref{bandstructure}(b) show the (empty) lowest 2D subband energy level in G2 is closer to the Fermi energy than in G1, predicting a lower turn-on threshold gate voltage for G2 than in G1. This is experimentally observed in Figure 1(c) of the main text. \textit{(ii)} At the same electron density $n_{2D}=2\times 10^{11}$ cm$^{-2}$ ($V_g > 0$, above turn-on threshold), Figures \ref{bandstructure}(c) and \ref{bandstructure}(d) show the 2DEG wavefunction peak in G2 is $\sim$\,3.5 nm further away than that of G1 from the Si dopants in the $n$-InSb layer, predicting a slightly higher mobility in G2 than in G1. This is experimentally observed in Figure 2(c) of the main text. \textit{(iii)} Figures \ref{bandstructure}(c) and \ref{bandstructure}(d) also show that the electric field across the 2DEG is more tilted in G1 than in G2, predicting a larger Rashba spin orbit coefficient in G1 than in G2. This is experimentally observed in Figure 4(b) of the main text.

\begin{figure}[t]
    \includegraphics[scale=0.85]{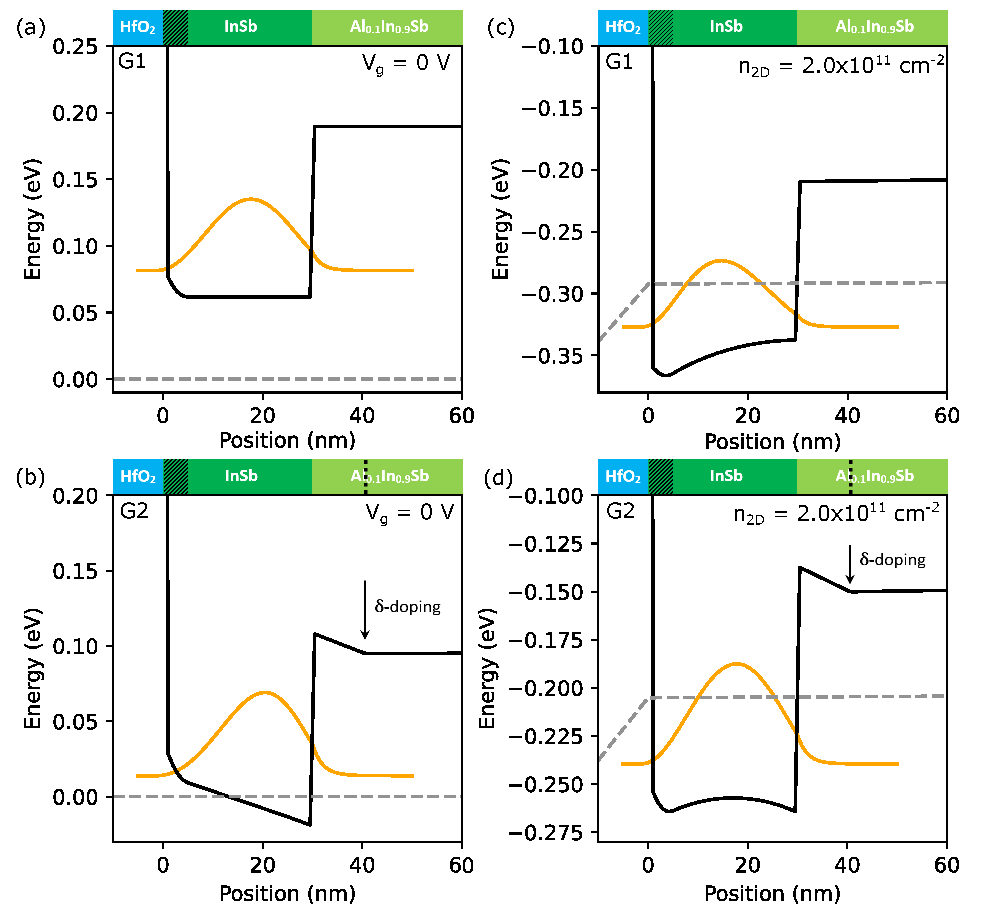}
    \caption{Calculated bandstructure profiles in the MBE growth direction of gated Hall bars fabricated from wafers G1 and G2. The corresponding MBE heterostructure is overlaid at the top of each panel, where ``0 nm'' corresponds to the HfO$_2$/InSb interface. The 5 nm thick $n$-InSb layer is indicated by the hatched area within the 30 nm InSb quantum well (dark green). For G2, the delta-doped layer in Al$_{0.1}$In$_{0.9}$Sb is indicated by a dotted line. Bandstructure profile of depleted 2DEGs at $V_g=0$ ($n_{2D}=0$) for: (a) G1 and (b) G2. Bandstructure profile of populated 2DEGs at $n_{2D}= 2 \times 10^{11}$ cm$^{-2}$ ($V_g > 0$) for: (c) G1 and (d) G2. In all four panels, the trapped charges associated with HfO$_2$, responsible for depleting the 2DEG after the dielectric deposition, are modeled by a delta-doped layer at the InSb/HfO$_2$ interface with a sheet density $N_{it}=1\times 10^{12}$ cm$^{-2}$, consistent with previously published reports.\cite{yi2015gate,Baik2017} The 2DEG wavefunction ($\Psi$) is represented by a slid orange line, the conduction band edge by a solid black line, and the Fermi level by a dashed grey line.}
    \label{bandstructure}
\end{figure}

\clearpage

\section{Fabrication methods}
\label{sec:fabrication}

The fabrication steps towards realization of all Hallbar devices discussed here and in the main text are presented in the following:

Samples are cleaned prior to lithography by sonication in acetone and subsequently propanol for 5 minutes each before a final blow dry with nitrogen. Mesa regions are defined with optical lithography using Shipley S1811 photoresist. The resist is spun at 5000 rpm for 60 seconds and baked at 120 $^\circ$C for 90 seconds. Following exposure, the photoresist is developed in MF319 developer for one minute. In order to ensure no unintentional thin film of photoresist remains in the exposed regions, samples are ashed in an oxygen plasma at 50 W for twenty seconds prior to wet etching to remove any residual photoresist in the exposed (off-mesa) regions. Wet etching proceeds with a ten second dip in buffered oxide etch (BOE) (1:10) to remove any native oxide on the surface of the sample caused by ashing and exposure to air. The mesa is etched with a solution of H$_2$O$_2$:H$_3$PO$_4$:C$_6$H$_8$O$_7$:H$_2$O mixed 3:4:9:44 by volume for approximately 30 seconds or until an etch depth of at least 100 nm has been reached. After etching, the photoresist etch mask is removed by sonication in acetone and isopropanol.

Optical lithography for definition of Ohmic contacts uses a bilayer resist recipe of MMA/Shipley. First the MMA (methyl methacrylate) is spun at 5000 rpm for 60 seconds and baked at 150 $^\circ$C for 5 minutes. Next the Shipley is spun in the same manner with a bake at 120 $^\circ$C for 90 seconds. Optical exposure and development of the sample in MF319 succesfully removes Shipley in regions where Ohmic contacts are to be formed. This exposure and development does not remove the MMA which protects the surface from being etched by the MF319 developer. MMA is subsequently removed by a fifteen minute exposure and development in a solution of isopropanol:H$_2$O at a 7:3 concentration. The now exposed surfaces are sulphur passivated in a solution of ammonium polysulfide (NH$_4$)$_2$S$_x$ for 20 minutes under illumination and at room temperature. Loading the sample into the deposition chamber proceeds immediately after passivation to minimize exposure to air. An angled 45$^\circ$ deposition of 20/60 nm of Ti/Au is performed in a thermal evaporator.

The 60 nm thick HfO$_2$ dielectric layer which isolates the top gate from the quantum well and Ohmic contacts in a gated Hallbar is deposited using atomic layer deposition at 150 $^\circ$C for 100 minutes; the dielectric breakdown field is $\sim$1.5 MV/cm at $T=1.6$ K. Following deposition, optical lithography with Shipley is used to define vias above the Ohmic contacts. The HfO$_2$ in the exposed vias is etched in BOE at a concentration of 1:10. Following etching, via resist is removed and processing proceeds with optical lithography of the top-gate and bond pads to metallic contacts. A bilayer of MMA/Shipley as discussed for the Ohmic contacts is again used and the Ti/Au (20/60 nm) top-gate and bond pads are similarly deposited in a thermal evaporator at an angle of 45$^\circ$.

\begin{figure}[h!]
    \includegraphics[scale=1.0]{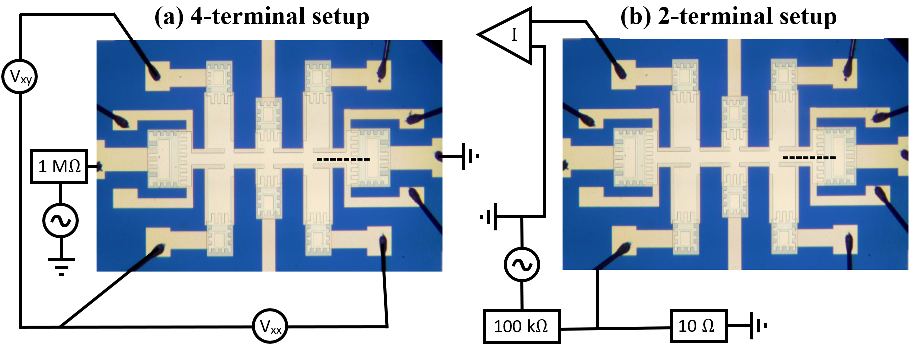}
    \caption{Electrical circuits for: (a) constant-current four-terminal setup with voltage preamplifiers ($\bigcirc$) for measuring $V_{xx}$ and $V_{xy}$, and (b) constant-voltage two-terminal setup with a current preamplifier ({\Large$\triangleright$}) for measuring differential conductance $G=dI/dV$. The ac oscillator ($\sim$) outputs a signal ranging from 10 mV to 1 Volt at low frequencies (10$-$20 Hz).}
    \label{fig:circuits}
\end{figure}

\section{Additional magnetotransport characterization}
\label{sec:magnetotransport}

Figure \ref{fig:circuits} shows the electrical circuits used in experiments. The typical ``constant'' ac voltage excitation in 2-terminal measurements was 100 \textmu V. The typical ``constant'' ac current in 4-terminal measurements was 100 nA for $T>1.5$ K and 10 nA for $T<100$ mK.

During Hall density and mobility constant-current 4-terminal measurements (Fig.\,1c and Fig.\,2c in the main text), the carrier density $n_{2D}$ was kept above 1$\times 10^{11}$ cm$^{-2}$ at all times. Otherwise, as the sample becomes more resistive, an increasingly significant fraction of the ac signal applied to the 1 M\textohm~resistor is dropped across the 2DEG. At pinch-off, the ac signal is entirely applied across the 2DEG rather than across the 1 M\textohm~resistor. Such voltages, which can be larger than the Fermi energy and even the confinement potential of the 2DEG in the InSb quantum well, can cause charging effects that last for the remainder of the cooldown (a thermal cycle to room temperature ``resets'' the device to its original characteristics).

Magnetotransport data of additional Hallbar devices fabricated in G1 and G2 is provided in Figure \ref{Additional_Magnetotransport}. Magnetotransport characteristics between Hallbars are quite reproducible indicating the quality of growth and fabrication. Furthermore, as discussed in the main text, there are no signs of parasitic parallel conduction or second subband occupation.

\begin{figure}[h!]
    \includegraphics[scale=0.9]{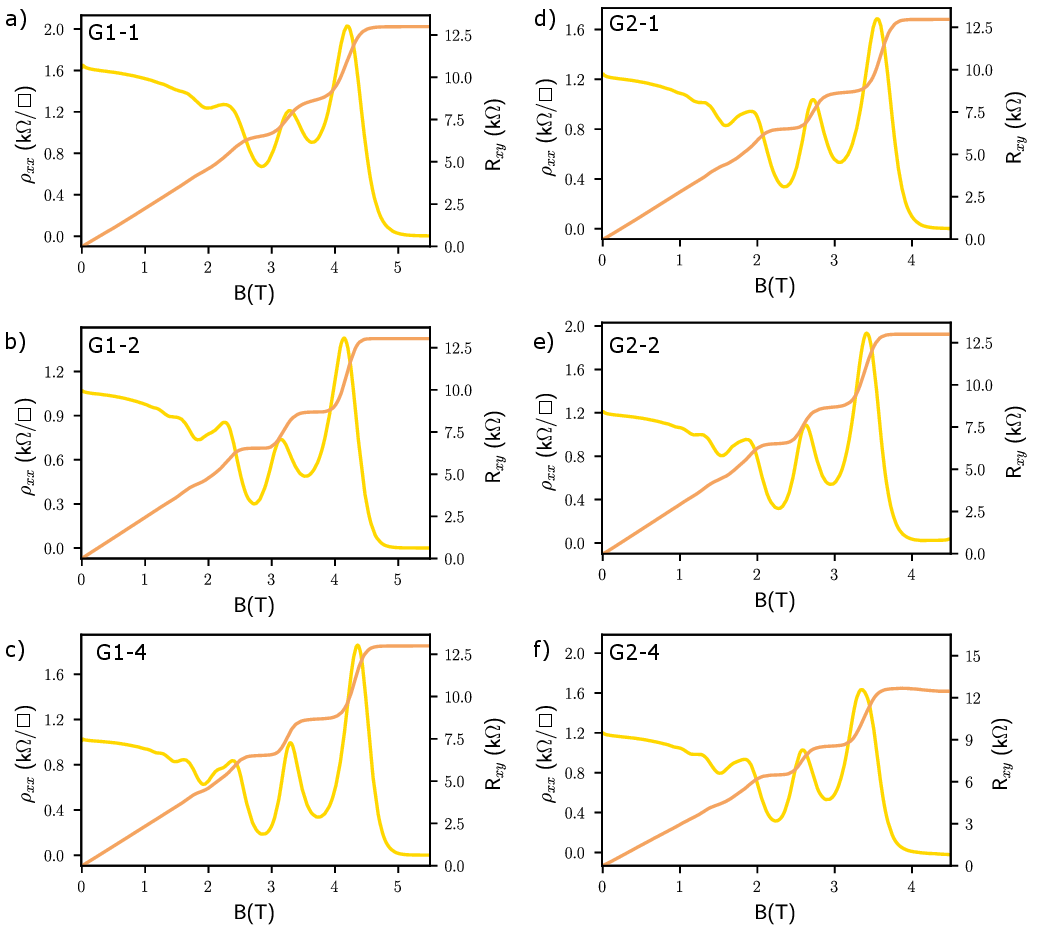}
    \caption{Longitudinal resistivity $\rho_{xx}$ (yellow traces) and Hall resistance $R_{xy}$ (orange traces) at $T=1.6$ K of additional samples in G1 near $2.7\times 10^{11}$ cm$^{-2}$ (a, b, c) and G2 near $2.2\times 10^{11}$ cm$^{-2}$ (d, e, f). We observe the oscillation in $\rho_{xx}$ corresponding to $\nu=2$ hit zero resistance, indicating the absence of parasitic conduction. Furthermore, the absence of a second oscillation frequency in all figures is indicative of single-subband occupation.}
    \label{Additional_Magnetotransport}
\end{figure}

\section{Effective Mass and quantum lifetime}
\label{sec:mass}

In Figure 3(a) of the main text, the amplitude of oscillations $\Delta\rho_{xx}$ is obtained by subtraction of a polynomial background resistance from $\rho_{xx}$. Data was taken at a fixed density of $3 \times 10^{11}$ cm$^{-2}$. The temperature dependent amplitude of SdH oscillations is described by a thermal damping term
\begin{equation}
    X(T) = \frac{2\pi^2k_BT/\hbar\omega_c}{\sinh(2\pi^2k_BT/\hbar\omega_c)}
\end{equation}
where $k_B$ is the Boltzmann constant, $T$ is temperature, and $\omega_c$ is the cyclotron frequency.\cite{ihn2009semiconductor} The effective mass, appearing in the cyclotron frequency, is determined from a least squares fitting of $X(T)$ to the temperature dependent amplitude of an oscillation at a given filling factor. A representative fit is presented in Figure 3b for the oscillation corresponding to $\nu = 8$ at $B = 1.56$ T.

The envelope of SdH oscillations is described by
\begin{equation}
    \Delta \rho_{xx} = 4\rho_{0}X(T)e^{-\pi/\omega_c\tau_q}
\end{equation}
where $\rho_0$ is the zero field resistivity, $\omega_c$ is the cyclotron frequency, and $X(T)$ is the thermal dampening term given previously.\cite{dingle1952some, coleridge1991small} At low enough temperatures where thermal damping can be neglected, the amplitude of oscillations is described by the Dingle term $e^{-\pi/\omega_c\tau_q}$. Using a so-called Dingle plot, as shown in Figure 3c in the main text, the quantum lifetime $\tau_q$ is given by the slope of $\ln{\Delta\rho_{xx}/4\rho_0X(T)}$ as a function of $1/B$.

\begin{table}[b]
    \caption{Material parameters used in the calculation, based on standard notation.\cite{Livneh2012}}
    \begin{ruledtabular}
    \begin{tabular}{ccc}
    \textbf{Parameter} & \textbf{InSb} & \textbf{Al$_{0.1}$In$_{0.9}$Sb} \\
    \hline
    $\mathrm{a}_{0}\left(\textrm{Å}\right)$ & 6.4794 & 6.44501 \\
    $m_{e}^{*}/m_{0}$ & 0.014 & 0.023 \\
    VBO (eV) & -0.053 & -0.1021 \\
    $E_{0}\left(\mathrm{eV}\right)$ & 0.237 & 0.4066 \\
    $\Delta_{0}\left(\mathrm{eV}\right)$ & 0.81 & 0.771 \\
    $E'_{0}\left(\mathrm{eV}\right)$ & 3.4 & 3.43 \\
    $\Delta'_{0}\left(\mathrm{eV}\right)$ & 0.4 & 0.39 \\
    $E_{\mathrm{P}}\left(\mathrm{eV}\right)$ & 22.8 & 22.3 \\
    $\gamma_{1}$ &  35.0800 & 21.2858 \\
    $\gamma_{2}$ &  15.6400 & 8.7492 \\
    $\gamma_{3}$ &  16.6306 & 9.7356 \\
    $a_{c}\left(\mathrm{eV}\right)$ &  -6.94 & -6.757 \\
    $a_{v}\left(\mathrm{eV}\right)$ &  -0.36 & -0.245 \\
    $b\left(\mathrm{eV}\right)$ &  -2 & -1.935 \\
    $c_{11}\left(\mathrm{Gdyne/cm^{2}}\right)$ &  684.7 & 704.2 \\
    $c_{12}\left(\mathrm{Gdyne/cm^{2}}\right)$ &  373.5 & 379.6
    \end{tabular}
    \end{ruledtabular}
    \label{tab:Parameter-set}
\end{table}

\section{8-BAND \lowercase{k}$\cdot$\lowercase{p} MODEL OF I\lowercase{n}S\lowercase{b}/A\lowercase{l}$_{0.1}$I\lowercase{n}$_{0.9}$S\lowercase{b} quantum well} \label{sec:kdotp}

The 8 band $\mathbf{k\cdot p}$ model of Livneh \textit{et al.} is
used with the parameters listed in Table \ref{tab:Parameter-set}
to estimate the effect of strain and quantum confinement on the in-plane
effective mass of InSb.\cite{Livneh2012,*[{ibid.~}] LivErratum2014} The model
has been shown in the past to give very good agreement with the band
gaps and absorption spectra of InAs/GaSb, InAs/AlSb and $\mathrm{InAs/InAs_{1-x}Sb_{x}}$
type II superlattices.\citep{KlipsteinJEM2014} Using Eq.\,C1 of
Ref.\,\onlinecite{Livneh2012}, $\gamma_{3}$ of the well material,
and the three Luttinger parameters, $\gamma_{1}$, $\gamma_{2}$ and
$\gamma_{3}$, of the barrier material, are calculated from $\gamma_{1}$
and $\gamma_{2}$ of the well, whose values we take from the work
of Lawaetz.\citep{Lawaetz1971} This reduces systematic errors introduced
when Luttinger parameters are taken from more than one source, and
is well suited to quantum wells (QWs) with a ternary barrier material
since it properly takes band bowing into account. The model also includes
interface parameters which are quite significant in the case of the
binary/binary T2SLs, but which are negligible in the present case
due to the low aluminium concentration in the barriers, whose major
constituent is the same as the binary quantum well material.

Figure \ref{fig:Comparison of bandstructures} compares the in-plane
band structures, $E\left(k_{||}\right)$, close to the band gap for
relaxed and strained InSb and for a strained $\mathrm{InSb/In_{0.9}Al_{0.1}Sb}$
superlattice with layer thicknesses of 93 ML / 70 ML (ML\,=\,monolayer\,$\approx3\textrm{Å}$),
where the strain of -0.53\% is provided by pseudomorphic growth on
relaxed $\mathrm{In_{0.9}Al_{0.1}Sb}$. Because the superlattice layers
are quite thick, there is negligible dispersion in the growth direction
for the conduction and valence bands shown in Fig.\,\ref{fig:Comparison of bandstructures},
so the superlattice can be viewed as a multiple quantum well (MQW),
where the in plane dispersion is the same as for a single QW.

When in-plane compressive strain is applied to bulk InSb, as shown
in Fig.\,\ref{fig:Comparison of bandstructures}, the hydrostatic
component tends to increase the band gap while the uniaxial component
tends to reduce it, by splitting the valence band so that the heavy-hole
(HH) is uppermost. Hence the band gap exhibits only a small net increase
and the HH in-plane dispersion shows a clear anti-crossing with the
light-hole (LH). Note that ``heavy'' and ``light'' refer to masses
in the growth- or \textit{z}-direction. This behaviour is reflected
in the QW, where the valence band edge is HH-like, with a series of
closely spaced HH sub-bands whose in-plane dispersions anti-cross
with the first LH sub-band. The main difference for the strained QW
is that it has a band gap that is 11.3\,meV (or 4.6\%) larger than
that of the strained InSb, due to the additional contribution of quantum
confinement. The band gap is 20.9\,meV (or 8.8\%) larger than that
of relaxed InSb.

\begin{figure}
  \includegraphics[scale=0.15]{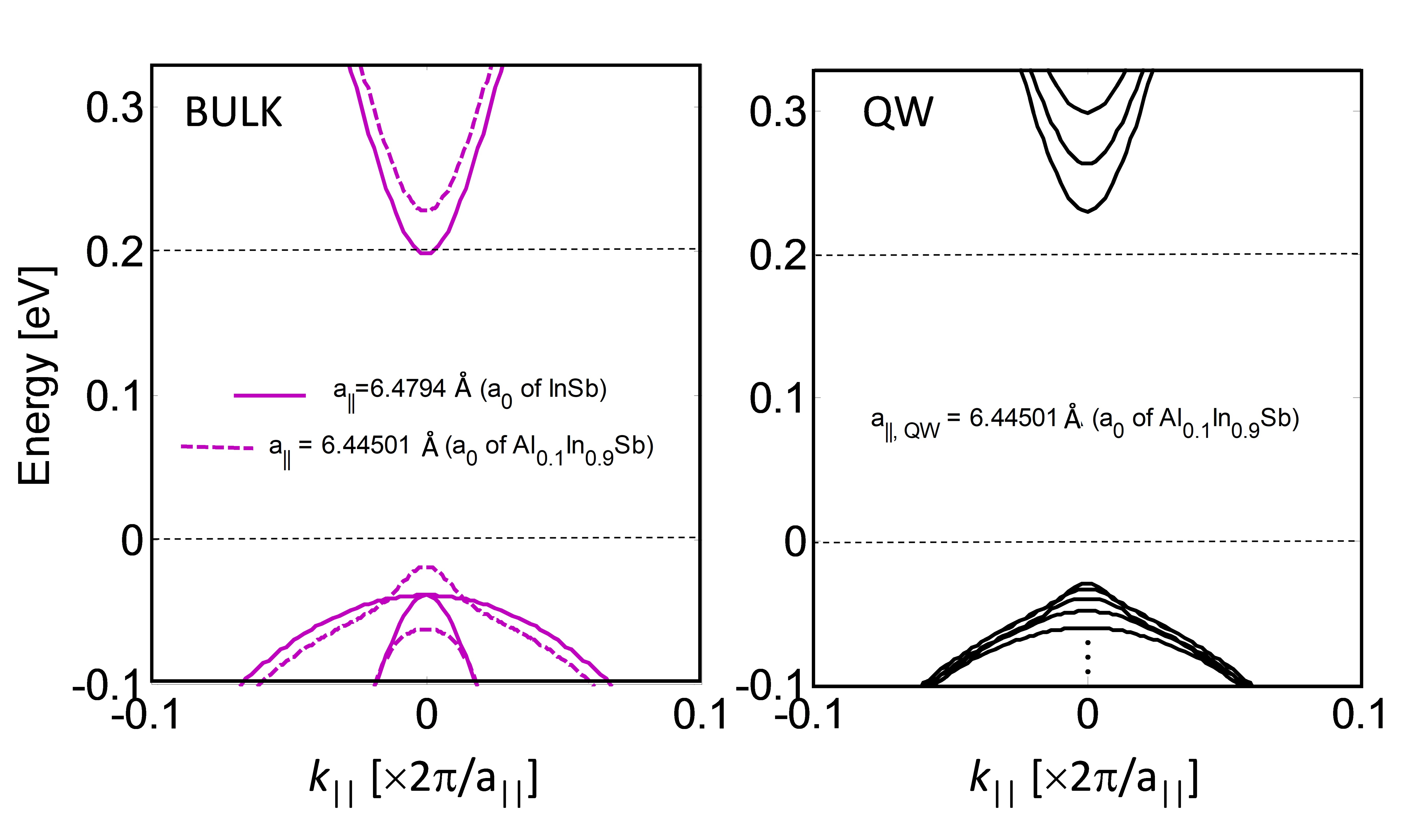}
  \caption{\label{fig:Comparison of bandstructures}Comparison of the band structure
  in the in-plane {[}100{]} direction: (left panel) for relaxed (solid line) and strained (dashed line) bulk InSb, and (right panel) for a strained 93\,ML\,/\,70\,ML MQW. For bulk InSb, the bands have been shifted in each case so that the edges of the conduction bands are identical with those of a 93\,ML\,/\,70\,ML MQW with the same in-plane lattice parameter. For the strained cases, the in-plane lattice parameter is that of relaxed $\mathrm{In_{0.9}Al_{0.1}Sb}$. Only the first 3 conduction sub-bands and the first 5 valence sub-bands are shown for the MQW. In the legends, {\small{}$\mathrm{a}_{0}$ is the cubic lattice parameter}.}
\end{figure}

In Figure \ref{fig:Eff mass}(a) the electron effective masses
and their ratio are shown for relaxed InSb and the strained QW. They
are found by applying the formula, $m^{*}=\tfrac{\hbar^{2}}{|\partial^{2}E/\partial k_{||}^{2}|}$,
to a sixth order polynomial that provides a very good fit to the $\mathbf{k\cdot p}$
dispersions in Fig.\,\ref{fig:Comparison of bandstructures} over the range, $|k_{||}|<0.022\times2\pi/\mathrm{a}_{\mathrm{InSb}}$.
Based on a simple two band QW Hamiltonian,\citep{Bernevig2006,Klipstein2021}
$H=A\left(\sigma_{x}k_{x}-\sigma_{y}k_{y}\right)+\sigma_{z}\left(\frac{E_{0}}{2}+Bk_{||}^{2}\right)+I_{0}Dk_{||}^{2}$,
the in-plane dispersion of the conduction band edge varies as $A^{2}k_{||}^{2}/E_{0}$
with an effective mass, $m^{*}=$$\hbar^{2}E_{0}/2A^{2}$ ($\sigma{}_{i}$
are the Pauli spin matrices, $I_{0}$ is the identity matri\textit{x,
A} is the electron-hole hybridization parameter, $E_{0}$ is the QW
band gap, and \textit{B}, \textit{D} represent interactions with remote
bands which are small and have been ignored). In the limit of infinite well width, $E_{0}\rightarrow E_{G}$,
where $E_{G}$ is the bulk band gap. Since \textit{A} scales inversely
with the lattice parameter,\citep{Gershoni93,Klipstein2021} the
electron band edge effective mass in the QW is decreased by 1.06\%
due to electron-hole hybridization, and increased by 8.8\% due to
the change in the band gap, giving an overall up shift of 7.7\%. This
is fairly close to the plotted value of 12.7\% in Fig.\,\ref{fig:Eff mass}(a),
suggesting that the two band model captures the essential physics
of the band edge effective mass fairly well, but there may be a small
additional contribution due to electron penetration of the barriers.

\begin{figure}
  \includegraphics[scale=0.13]{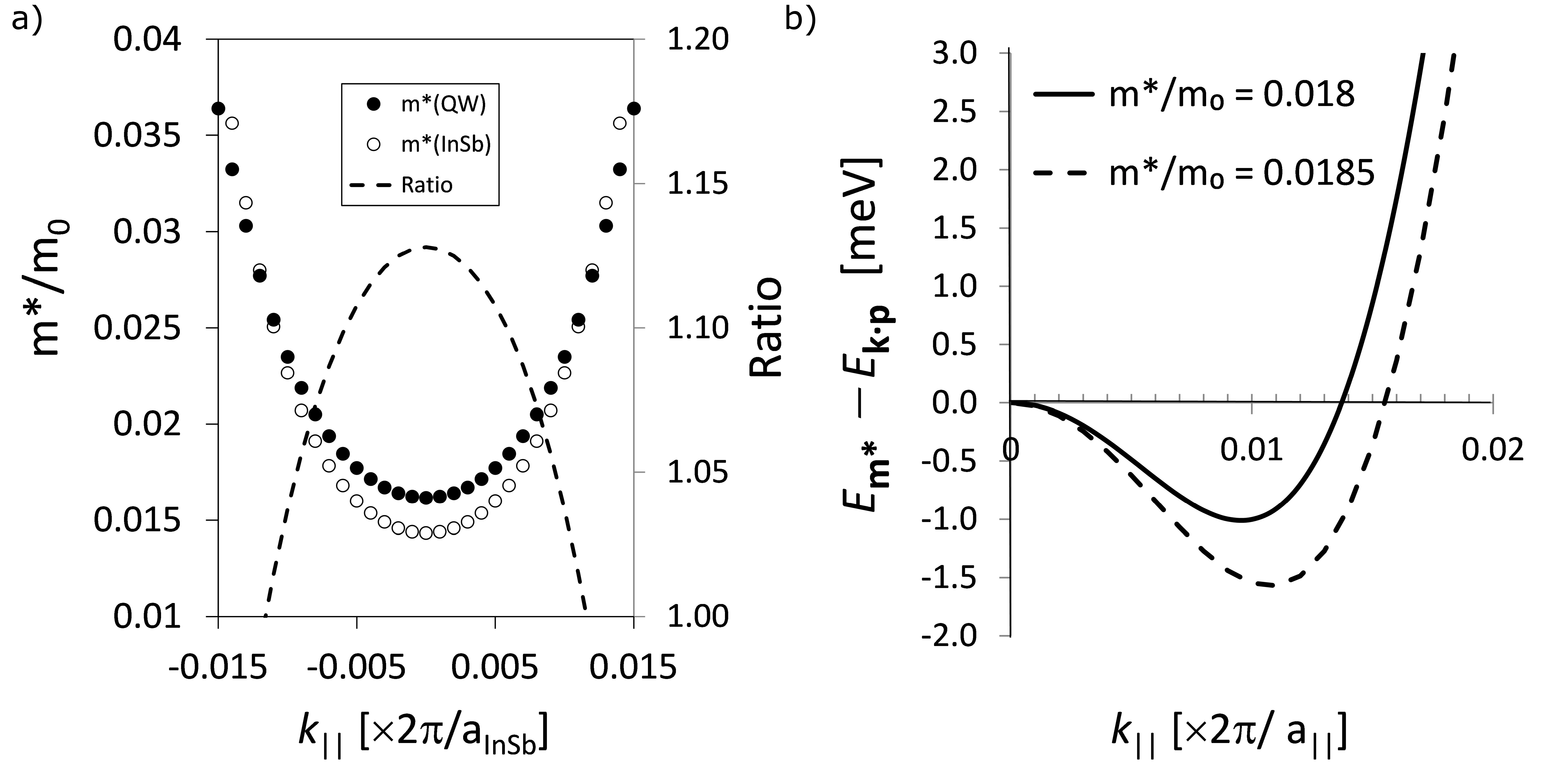}
  \caption{\label{fig:Eff mass}(a)\,Calculated band curvature effective masses
  in terms of the free electron value, $m_{0}$, and their ratio for
  relaxed InSb and the strained MQW ($\mathrm{a_{InSb}=a_{0}}$ of InSb)
  (b) Difference between a parabollic dispersion and the $\mathbf{k\cdot p}$
  dispersion of the QW shown in Fig.\,\ref{fig:Comparison of bandstructures},
  for different values of the effective mass, $m^{*}$.}
\end{figure}

The rapid increase of the band curvature effective mass with wave
vector in Fig.\,\ref{fig:Eff mass}(a) shows that strong non-parabolicity
exists in the conduction band of both bulk InSb and the QW. For the
2DEG density of $3\times10^{11}\thinspace\mathrm{cm^{-2}}$ in Fig.
3(a) of this letter, the electron Fermi wave vector of $k_{\mathrm{F}}=0.014\times2\pi/\mathrm{a}_{\mathrm{InSb}}$, corresponds to a band curvature effective mass in the QW of 0.033\,$m_{0}$.
This value does not agree with 0.019\,$m_{0}$ measured at \textit{B}\,=\,1.56\,T
in Fig. 3(b), because the magneto-transport assumes a parabollic model,
whose mass is used to determine the Landau energies: $E_{N\uparrow,\downarrow}=(N+\frac{1}{2})\frac{\hbar eB}{m^{*}}\pm\frac{1}{2}g\mu_{B}B$.
This parabolic mass value is fitted to the temperature dependent
amplitude of the SdH oscillations, where electrons are thermally excited
from nearly filled to nearly empty Landau levels.\citep{Singleton2001}
Therefore we need to find a parabolic dispersion that intersects
the $\mathbf{k\cdot p}$ dispersion close to the Fermi wave vector.
At this wave vector, the number of states within the zero field $\mathbf{k\cdot p}$
Fermi circle matches the number of filled Landau states. Figure \ref{fig:Eff mass}(b)
shows that the difference between the parabolic and $\mathbf{k\cdot p}$
dispersion energies vanishes at $k_F = 0.0137 \times 2\pi/a_{||}$ when the parabolic mass is 0.018\,$m_{0}$.\footnote{Note that $\mathrm{a}_{||}$ is used here for the QW while $\mathrm{a}_{\mathrm{InSb}}$ at $k_{\mathrm{F}}=0.0137\times2\pi/\mathrm{a}_{\mathrm{||}}$ was used earlier, but the difference is small enough to yield the same prefactor close to 0.014.} If we add the number of states in the next (empty) Landau level at 1.56\,T for both spin
directions, to take into account their role in the temperature dependence
of the SdH oscillations, the wave vector for the circle that includes
all of these states increases to $k_{\mathrm{F}}^{*}=0.0157\times2\pi/\mathrm{a}_{\mathrm{||}}$.
Figure \ref{fig:Eff mass}(b) shows that the effective mass corresponding
to this circle increases to 0.0185\,$m_{0}$. Thus an average value
close to 0.0183\,$m_{0}$, is expected to correspond to the measured
SdH mass. Since the latter was found to be 0.019\,$m_{0}$, the agreement
between the $\mathbf{k\cdot p}$ model and experiment appears to be
quite reasonable.

\section{Coincidence Measurement and \lowercase{g}-factor}
\label{sec:gfactor}

In Figures \ref{Coincidence}(a) and \ref{Coincidence}(b) the longitudinal resistivity $\rho_{xx}$ as a function of the perpendicular magnetic field $B_\perp$ for different tilt angles $\theta$ (see inset Fig.S1(b)) is shown for G1 at densities corresponding to (a) 2.8$\times10^{11}$ cm$^{-2}$ and (b) 3.6$\times10^{11}$ cm$^{-2}$. At $\theta = 0^{\circ}$, we observe the onset of spin splitting at $\nu = 5$ around 2 T followed by both even and odd filling factors corresponding to $\nu = 4, 3, 2$ at higher fields. As the tilt angle is increased, the width of the minima in $\rho_{xx}$ decreases for even integer filling factors ($\nu = 2, 4$) and increases for odd integer filling factors ($\nu = 3, 5$).
Eventually, peaks will coalesce at even integer filling factors as minima in $\rho_{xx}$ at odd integer filling factors approach their largest widths. The coalescing of peaks in this case corresponds to the crossing of spin split Landau levels of different spin polarizations and is used to determine the effective g-factor.

\begin{figure}
    \centering
    \includegraphics[scale=0.95]{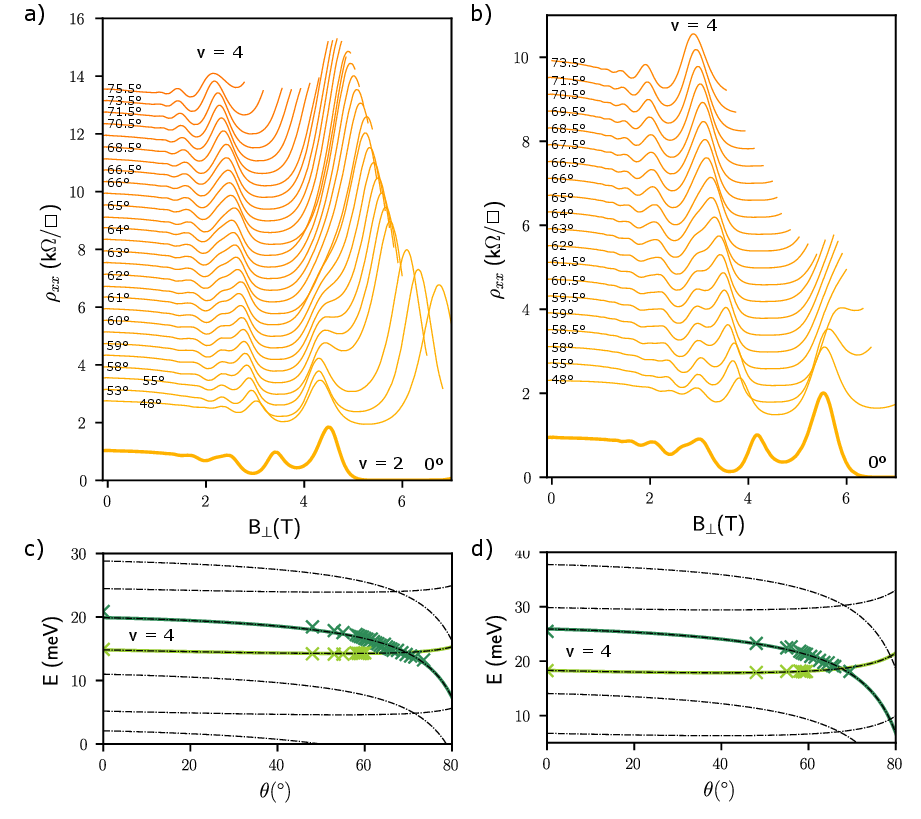}
    \caption{\label{Coincidence} Coincidence measurement. (a) Longitudinal resistivity versus magnetic field at (a) $V_g=0.75$ V and (b) $V_g=0.85$ V is taken at various tilt angles $\theta$ with respect to normal vector of the sample surface. The perpendicular field values $B_{\perp}$ of peaks in resistivity surround $\nu=4$ in (a) and (b) are plotted versus tilt angle $\theta$ in (c) and (d) respectively.}
\end{figure}

Figures \ref{Coincidence}(c) and \ref{Coincidence}(d) show the $B_\perp$ values of the peaks in the SdH oscillations shown in Figs.\,\ref{Coincidence}(a) and \ref{Coincidence}(b) respectively as a function of tilt angle $\theta$. Peaks corresponding to the observed crossing at $\nu = 4$ in Figs.\,\ref{Coincidence}(a) and \ref{Coincidence}(b) are presented in \ref{Coincidence}(c) and \ref{Coincidence}(d) respectively. The evolution of peaks in $\rho_{xx}$ as a function of $\theta$ is described by the evolution of the Landau level energy spacing described by $E_N = \hbar\omega_c(\theta)(N + 1/2) \pm \frac{1}{2} g^*\mu_B B_{\text{tot}}$ where $\hbar$ is the reduced Plank's constant, $\omega_c (\theta)=eB_\perp(\theta)/m^*$ is the cyclotron frequency, $N=0,1,2,...$ is an integer, $g^*$ is the effective g-factor, $\mu_B$ is the Bohr magneton, and $B_{\text{tot}}$ is the total magnetic field.

All scans were taken at two fixed gate voltages ($V_g=0.75$ V and $V_g=0.85$ V), which in this case did not correspond to a fixed density. Operating the piezo-electric rotator stage
over the duration of the experiment was observed to change the relation of $n_{2D}(V_g)$. This particular sample, G1-3, had been otherwise very stable in many cooldowns in two other cryostats. For example, the stable pinch-off curves in Figure 1(c), the stable Landau fan in Figure 2(b), and the temperature dependence of the WAL peak in Section VII of the Supplementary were all performed on sample G1-3, with density remaining stable and reproducible for weeks at a time. We thus cannot explain the density instability between scans at different tilt angles (during the scan, the density remains stable throughout), other than perhaps due to the heat pulse generated while the rotator to a different angle $\theta$ moved between scans. In any case, having measured the carrier density of each $B_\perp$ scan via the Hall effect, we modeled the density-driven change in the Landau level energy for each scan by using $\omega_c = eB_{\perp}/m^*$ and $\nu = hn_{2D}/eB_\perp$ for a given filling factor $\nu$. With this correction, a best fit of the spin split energy levels (solid lines) to the data (crosses) yielded an effective g-factor of $33\pm 2$ at $\nu=4$ in (c) and $41 \pm 2$ at $\nu=4$ in (d).

\begin{figure}
    \centering
    \includegraphics[scale=0.75]{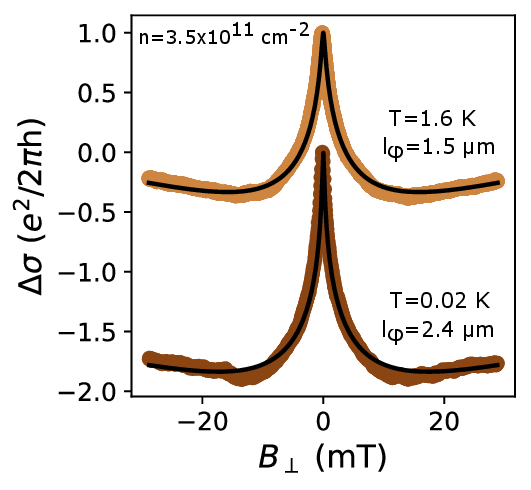}
    \caption{\label{WAL_Temp} Weak antilocalization measurements of sample G1-3 taken at $T=20$ mK and $T=1.6$ K. The results of fits to the HLN model (black lines) are reported in figure 4c of the main text. }
\end{figure}

\section{Weak Anti-localization and spin-orbit interactions}
\label{sec:WAL}

In systems with strong spin-orbit interaction, the longitudinal conductivity in small magnetic fields exhibits a pronounced peak at B = 0 due to the suppression of coherent backscattering. In our measurement, the longitudinal conductivity $\sigma_{xx}$ (B) is determined from simultaneous measurements of the longitudinal and transverse resistances. As shown in figure 4a of the main text, the conductivity correction $\Delta\sigma_{xx} (B) = \sigma_{xx} (B) - \sigma_{xx} (0)$ exhibits a peak in both G1 and G2 for various densities. The strength of SOI is quantified from fits of the conductivity correction to the Hikami Larkin Nagaoka model.\cite{hikami1980spin} The conductivity correction of the HLN models reads:
\begin{align}
        \Delta\sigma_{xx} (B) ={}& \frac{e^2}{2\pi^2\hbar}\bigg[\Psi\big(\frac{1}{2} + \frac{H_{\phi}}{B} + \frac{H_{so}}{B}\big) + \frac{1}{2}\Psi(\frac{1}{2} + \frac{H_{\phi}}{B}+ \frac{2H_{so}}{B})\\
        & - \frac{1}{2}\Psi(\frac{1}{2} + \frac{H_{\phi}}{B}) - \ln(\frac{H_{\phi} + H_{so}}{B}) - \frac{1}{2}\ln(\frac{H_{\phi} + 2H_{so}}{B}) \nonumber \\
        & + \frac{1}{2}\ln(\frac{H_{\phi}}{B})\bigg]. \nonumber
\end{align}
The fit parameters H$_{\phi}$ and H$_{SO}$ correspond respectively to the phase coherence and spin-orbit effective fields and $\Psi$ is the Digamma function. The fit parameters can be converted to their corresponding lengths using $l_{\phi} = \frac{\hbar}{4eH_{\phi}}$ and $l_{so} = \sqrt{\tau_{so}D}$ where $D$ is the diffusion constant. Figure \ref{WAL_Temp} displays fits of the HLN model to the conductivity correction measured in sample G1-3. As reported in the main text and shown here, the phase coherence reached 2.4 \textmu m at 20 mK.

\begin{table}[t]
    \begin{ruledtabular}
    \begin{tabular}{ccccc}
    \textbf{Source} & \textbf{InSb QW width} & \textbf{2DEG depth} & $\alpha_{so}$ & $n_{2D}$ \\
    ~ & \textbf{(nm)} & \textbf{(nm)} & \textbf{(meV$\cdot\text{\AA}$)} & \textbf{(10$^{11}$ cm$^{-2}$)} \vspace{0.5mm}\\
    \hline
    Reference \onlinecite{gilbertson2009zero} & 30 & 50 & 130 $-$ 150 & 3.2 $-$ 3.3 \\
    Reference \onlinecite{khodaparast2004spectroscopy} & 30 & 160 & 130 & 1.9 $-$ 2.7 \\
    Our work~ & 30 & 0 & 80 $-$ 110 & 2.7 $-$ 4.6 \\
    Reference \onlinecite{lei2022arxiv} & 21.5 & 8 & 34 $-$ 91 & 1.9 $-$ 4.9 \\
    Reference \onlinecite{lei2022high} & 21 & 50 & 27 $-$ 31 & 1.4 $-$ 2.2 \\
    Reference \onlinecite{kallaher2010} & 25 & 163 & 30 & 5.5 \\
    \end{tabular}
    \end{ruledtabular}
    \caption{List of reported Rashba spin orbit coefficient $\alpha_{so}$ in literature in InSb/InAlSb quantum wells. In all cases shown here, the higher $\alpha_{so}$ corresponds to a larger electron density $n_{2D}$.}
    \label{TableS2}
\end{table}

Table \ref{TableS2} compares the values of $\alpha_{so}$ obtained from the WAL fits, and compares them to the literature values obtained in undoped and modulation-doped InSb 2DEGs.\cite{gilbertson2009zero,khodaparast2004spectroscopy,lei2022arxiv,lei2022high,kallaher2010} Our $\alpha_{so}$ reaches a maximum of nearly 110 meV$\cdot\text{\AA}$ at $n_{2D}=4.6\times 10^{11}$ cm$^{-2}$ in wafer G1, and is among the highest values reported. For surface or near-surface InSb quantum wells, we are reporting the highest value.

\providecommand{\noopsort}[1]{}\providecommand{\singleletter}[1]{#1}